\documentclass[aps,prl,twocolumn,reprint,superscriptaddress,nobibnotes]{revtex4}
\usepackage{graphicx,xcolor}
\usepackage{amsmath}
\usepackage{amssymb}
\usepackage{colordvi}
\usepackage{mathrsfs}
\usepackage{bm}
\usepackage{verbatim}
\usepackage{dcolumn}
\usepackage{epsfig}
\usepackage{subfigure}

\usepackage{mathtools}
\usepackage{braket}

\usepackage{hyperref}

\newcommand{\pd}[2]{\frac{\partial #1}{\partial #2}}

\begin{document}
\title{Extrinsic Mechanisms of Phonon Magnetic Moment}

%{
%\makeatletter
%\def\frontmatter@thefootnote{%
%\altaffilletter@sw{\@fnsymbol}{\@fnsymbol}{\csname c@\@mpfn\endcsname}%
%}%
%\makeatother

\author{Rui Xue}
\affiliation{International Centre for Quantum Design of Functional Materials, CAS Key Laboratory of Strongly-Coupled Quantum Matter Physics, and Department of Physics, University of Science and Technology of China, Hefei, Anhui 230026, China}
\affiliation{Hefei National Laboratory, University of Science and Technology of China, Hefei 230088, China}
\author{Zhenhua Qiao}
\email[Correspondence author:~]{qiao@ustc.edu.cn}
\affiliation{International Centre for Quantum Design of Functional Materials, CAS Key Laboratory of Strongly-Coupled Quantum Matter Physics, and Department of Physics, University of Science and Technology of China, Hefei, Anhui 230026, China}
\affiliation{Hefei National Laboratory, University of Science and Technology of China, Hefei 230088, China}
\author{Yang Gao}
\email[Correspondence author:~]{ygao87@ustc.edu.cn}
\affiliation{International Centre for Quantum Design of Functional Materials, CAS Key Laboratory of Strongly-Coupled Quantum Matter Physics, and Department of Physics, University of Science and Technology of China, Hefei, Anhui 230026, China}
\affiliation{Hefei National Laboratory, University of Science and Technology of China, Hefei 230088, China}
\author{Qian Niu}
\affiliation{International Centre for Quantum Design of Functional Materials, CAS Key Laboratory of Strongly-Coupled Quantum Matter Physics, and Department of Physics, University of Science and Technology of China, Hefei, Anhui 230026, China}
%\affiliation{Hefei National Laboratory, University of Science and Technology of China, Hefei 230088, China}
%\author{Qian Niu}
%\affiliation{CAS Key Laboratory of Strongly-Coupled Quantum Matter Physics, 
%	and 
%	Department of Physics, University of Science and Technology of China, 
%	Hefei, 
%	Anhui 230026, China}

\date{\today{}}

\begin{abstract}
  {We develop a general formalism of phonon magnetic moment by including the relaxation processes. We then identify the skew-scattering and side-jump contributions to the phonon magnetic moment originating from the non-adiabaticity, both of which are related to the nonlocal phonon Berry curvature and are in close analogy to those in the electronic Hall effect. All contributions of the phonon magnetic moment are exemplified  in a honeycomb lattice, showing that the extrinsic contribution can be as important as the intrinsic one and that the resulting phonon angular momentum varies significantly across the Brillouin zone.  Our work offers a systematic framework of the phonon chirality and paves the way of tuning the phonon magnetic moment through the non-adiabaticity.}
\end{abstract}
\maketitle

Recent years have seen a surge of interest in phonon chirality~\cite{Zhang2014,Zhu2018,Zhang2018,Miao2018,Nomura2019,Saito2019,Zhang2020,Cheng2020,He2020,Li2021,Ren2021,Hu2021,Bistoni2021,Saparov2022,Zhang2022,Juraschek2022,Zabalo2022,Zhang2023,Ueda2023,Gao2023,Bonini2023,Lujan2024,Tang2024,Mustafa2025}. As a unique degree of freedom characterizing the circular motion of phonons, phonon chirality  can couple to the spin and valley degree of freedom of electrons and hence plays important roles in various quantum phenomena, such as exciton formation and relaxation~\cite{Chen2020,He2020,Lujan2024}, magnetic dynamics~\cite{Zhang2019,Wang2024}, unconventional superconductivity~\cite{Gao2023}, etc.. As phonon modes can generally carry Born effective charges, their circular motion naturally leads to a nontrivial magnetic moment, which not only directly couples to the external magnetic field through the Zeeman effect~\cite{Nomura2019,Cheng2020,Juraschek2022} but also is responsible for the phonon Hall effect~\cite{Strohm2005,Sheng2006,Inyushkin2007,Kagan2008,Wang2009,Zhang2010,Juraschek2022}, a fundamental response of phonons. Under the adiabatic Born-Oppenheimer approximation, the phonon magnetic moment has been related to the Berry curvature of electrons in the parameter space characterizing the motion of nucleus, and hence favours topological systems~\cite{Saito2019,Cheng2020,Ren2021,Hu2021,Bistoni2021,Saparov2022,Bonini2023}.

As successful as it is, the current formulation of the phonon magnetic moment does not take into account the relaxation processes, which is essential in understanding the chiral response of electrons~\cite{Nagaosa2010,Sinitsyn2006,Kaplan2020,Kaplan2023,Raimondi2011,Xiao2019}, such as the anomalous Hall effect, spin Hall effect, etc.. Its basis, i.e., the Born-Oppenheimer approximation, relies on the fact that the electron mass is much smaller than that of atoms, such that the characteristic energy scale of the electronic motion to be much larger than that of the atomic motion. However, the relaxation processes introduce non-Hermiticity by the addition of extra energy scales to the phononic and electronic systems, originating from either externally controlled gain and loss or the internal correlation effects~\cite{Shen2018,Chen2018,Yao2018,Yokomizo2019,Pan2020,Kaplan2020,Michishita2020,Nagai2020,Wang2022,Geng2023,Kaplan2023,Yu2024,Kozii2024,Yang2024}. As these energy scales do not necessarily fulfill the requirement of the Born-Oppenheimer approximation, it is thus highly desirable to unravel their effect on the phonon magnetic moment. 

In this work, we present a microscopic theory of phonon magnetic moment using the time-dependent density-matrix perturbation theory. The relaxation processes are added through effective lifetime parameters to the phonon frequency and electronic Hamiltonian. A straightforward modification to the phonon magnetic moment is that the electron band gap is renormalized by the phonon frequency. As a result, the leading order correction is proportional to the ratio of the phonon frequency and the electron energy gap, as anticipated from the adiabatic  approximation. 

Strikingly, we also identify the skew-scattering and side-jump contributions to the phonon magnetic moment. Both vanish when the adiabaticity is restored in the relaxation processes and depend on the ratio of different lifetimes otherwise.  Microscopically, they originate from the molecular Berry curvature which characterizes the strength of the chiral-phonon-chiral-electron coupling. More precisely, the skew-scattering and side-jump contribution are due to the asymmetric scattering rate and the change of coordinate in the phonon-electron interaction, respectively, in analogy to those in the electronic Hall effect~\cite{Nagaosa2010}.  Our theory is exemplified in the honeycomb lattice, showing that the extrinsic contribution can be as important as the intrinsic one and that it can induce a sizable phonon angular momentum.

\textit{Microscopic theory.---} The phonon magnetic moment $\bm m$ manifests as a Zeeman-type energy shift under a magnetic field $\bm B$: $\hat{H}_{Z}=-\bm B\cdot \bm m$. It is proportional to the circular atomic motion~\cite{Zhang2010,Saito2019}:
\begin{equation}\label{eq_ram}
	m_i=\lambda_{ij} (\bm u^a\times \dot{\bm u}^a)_j\,,
\end{equation}
where $\lambda_{ij}$ is the response coefficient and $\bm u^a$ is the atomic displacement at lattice site $a$. Therefore, $\hat{H}_Z$ is the Raman term in the phonon Hamiltonian and is essential for the phonon Hall effect.

To derive $\bm m$ or equivalently $\lambda_{ij}$, we observe that the Raman term is quadratic in $\bm u$ and linear in $\bm B$. We then consider the electronic Hamiltonian involving the electron-phonon coupling and the magnetic field, and expand it up to the order of interest. The result reads~(for simplicity, we set $e=\hbar=1$)
\begin{align}\label{eq_exp}
	\hat{H}=&\hat{H}_{0}+\hat{H}_B+\hat{H}_{ep}+\hat{H}_{mix}\,,
\end{align}
where $H_{0}$ is the bare electronic Hamiltonian, $\hat{H}_B=\hat{\bm v}\cdot\bm A$ with $\bm A$ being the vector potential due to external magnetic field, $\hat{H}_{ep}=\hat{F}_i^a u_i^a+\hat{F}_{ij}^{ab} u_i^a u_j^b$ describes the electron-phonon coupling, $\hat{F}_i^a=\partial{\hat{H}_0}/\partial R_i^a$~($ R_i^a$ is the equilibrium atomic position at lattice site $a$), $\hat{F}_{ij}^{ab}=(1/2)\partial^2 \hat{H}_0/\partial R_i^a \partial R_j^b$, $\hat{H}_{mix}=\hat{F}_{i,j}^a u_{i}^aA_{j}+\hat{F}_{ij,k}^{ab}u_{i}^a u_j^b A_{k}$ describes the coupling between the electron-phonon interaction and the magnetic field, $\hat{F}_{i,j}^{a}=\partial \hat{v}_j/\partial R_i^a$, and $\hat{F}_{ij,k}^{ab}=(1/2)\partial^2 \hat{v}_k/\partial R_i^a \partial R_j^b$. Here and hereafter, Einstein summation convention is implied for repeated indices. Without loss of generality, we will consider the part of electron-phonon coupling with $a=b$, i.e., the onsite coupling. The generalization to $a\neq b$ is straightforward.

To proceed, we consider a single phonon mode: $u_i\rightarrow u_i(\omega,\bm q)e^{-i\omega t+i\bm q\cdot \bm r}$. Then $\bm u\times \dot{\bm u}\rightarrow \bm u(\omega,\bm q)\times \bm u(\omega,\bm q)^\star$. The magnetic field can be included by allowing the vector potential $\bm A$ varies in the form of $\bm A=\bm A(\bm q_0) e^{i\bm q_0\cdot \bm r}$ so that $\bm B=i\bm q_0\times \bm A$. 
%\begin{align}\label{eq_lam}
%\hat{H}_Z=\frac{1}{2}\omega\lambda_{ij}(\omega,\bm q) [\bm u(\omega,\bm q)\times \bm u(-\omega,-\bm q)]_i B_j\,.
%\end{align}
The coefficient $\lambda_{ij}(\omega, \bm q)$ can then be derived by integrating out the electronic degree of freedom in the full Hamiltonian in Eq.~\eqref{eq_exp}. We note that the second terms in both $\hat{H}_{ep}$ and $\hat{H}_{mix}$ do not contribute to $\lambda_{ij}$, as they are symmetric with respect to $u_i$ and $u_j$. For the other terms in Eq.~\eqref{eq_exp}, we seek the following response functions
\begin{align}\label{eq_res}
\langle v_k e^{i\bm q_0 \bm r}\rangle&=\theta^1_{kij}(\omega, \bm q,\bm q_0) u_i(\omega, \bm q) u_j(-\omega,-\bm q-\bm q_0)\,,\notag\\
\langle \hat{F}_i^a e^{i\bm q\cdot \bm R^a}\rangle&=\theta^2_{kij}(\omega,\bm q,\bm q_0) u_j(-\omega,-\bm q-\bm q_0) A_k(\bm q_0)\,,\notag\\
\langle \hat{F}_{i,k}^a e^{i(\bm q+\bm q_0)\cdot \bm R^a}\rangle&=\theta^3_{kij}(\omega,\bm q,\bm q_0) u_j(-\omega,-\bm q-\bm q_0)\,.
\end{align}
Combining with the Zeeman energy shift and Eq.~\eqref{eq_ram}, $\lambda_{ij}$ can then be expressed as follows~\cite{suppl}:
\begin{widetext}
\begin{align}\label{eq_lam2}
\lambda_{ij}(\omega,\bm q)&=\frac{1}{4}\epsilon_{j\ell k}\epsilon_{ik_1k_2}\lim_{\bm q_0\rightarrow 0} \partial_{q_{0\ell}}\partial_{\omega}\sum_{i=1}^{3}[\theta^i_{k k_1 k_2}(\omega, \bm q,\bm q_0) +\theta^i_{kk_2 k_1}(-\omega, -\bm q,\bm q_0)]\,.
\end{align}
\end{widetext}

The response functions in Eq.~\eqref{eq_res} can be derived using the density-matrix perturbation theory. The density matrix $\hat{\rho}$ satisfies the following equation~\cite{Michishita2020}:
\begin{align}\label{eq_dyn}
	\dot{\hat{\rho}} = -i (\hat{\rho}\hat{H}-\hat{H}^\dagger \hat{\rho})\,.
\end{align}
Originally $\hat{H}=\hat{H}^\dagger$ as in Eq.~\eqref{eq_res}. To include the relaxation processes, we add non-Hermitian self-energy part to $\hat{H}$ which manifests as an imaginary frequency in the phonon degree of freedom, i.e., $\omega\rightarrow \omega+i\eta_p$, and an imaginary self-energy $\Sigma$ in the electron degree of freedom. Without loss of generality, we further assume that both $\eta_p$ and $\Sigma$ are constant, and that $\Sigma$ is band-diagonal~\cite{Kaplan2020}, i.e., $\Sigma=i\eta_m\delta_{mn}$ with $m$ being band index.

We then solve Eq.~\eqref{eq_dyn} order-by-order by taking $u(\omega, \bm q)$ and $\bm A(\bm q_0)$ as small quantities, yielding $\hat{\rho}=\hat{\rho}_0+\hat{\rho}_1+\hat{\rho}_2+\cdots$. The response function $\theta^1_{kij}$ and $\theta^2_{kij}$ can be derived using $\hat{\rho}_2$ and $\theta^3_{kij}$ using $\hat{\rho}_1$. The phonon magnetic moment, or $\lambda_{ij}$ can then be obtained using Eq.~\eqref{eq_lam2}. 

Interestinly, we identify two distinct types of terms in $\lambda_{ij}$:
\begin{align}
\lambda_{ij}(\omega,\bm q)=\lambda_{ij}^{A}(\omega,\bm q)+\lambda_{ij}^{NA}(\omega,\bm q)\,.
\end{align}
  The first part $\lambda_{ij}^A$ can be smoothly reduced to the phonon magnetic moment under the Born-Oppenheimer approximation in semiconductors and insulators~\cite{suppl}. Its leading order correction is in the form of $\omega/\varepsilon_0$ where $\varepsilon_0$ is the energy scale of the electronic band, in accordance with the adiabatic approximation.

In sharp contrast, the second term $\lambda_{ij}^{NA}$ originates from non-adiabaticity and reads as:
\begin{widetext}
\begin{align}\label{eq_na}
\lambda_{ij}^{NA}&=-\frac{1}{2}\sum_{n_1n_3}\int d\bm k[\delta_{n_1}^{n_3}  (m^e_j)_{n_1\bm k}G_1(\omega,\bm q)(\Omega_i)_{n_1n_1}^{n_2}(\bm q)+\epsilon_{jj_1j_2}\delta_{n_1}^{n_3} (v_{j_2})_{n_1\bm k}G_2(\omega,\bm q) T(\bm q)-(\omega,\bm q\rightarrow -\omega, -\bm q)]\,,
\end{align}
\end{widetext}
where $\delta_{n_1}^{n_3}=\eta_{n_3}/(2\eta_p+\eta_{n_1})$, $\bm m^e_{n\bm k}=1/2 {\rm Im}\langle \bm \partial n\bm k| \times \hat{\bm v}|n\bm k\rangle$ is the electronic orbital magnetic moment in band $n$, $|n\bm k\rangle$ is the periodic part of the electronic Bloch wave function with energy $\varepsilon_{n\bm k}$ and distribution function $f_{n\bm k}$,  $(v_{j_2})_{n_1\bm k}=\langle n_1\bm k|\hat{v}_{j_2}|n_1\bm k\rangle$,
\begin{align}
G_1(\omega, \bm q)&=\frac{\Delta f_{n_1\bm k,n_2\bm k-\bm q} \Delta \varepsilon_{n_1\bm k,n_2\bm k-\bm q}^2}{(-\omega+\Delta \varepsilon_{n_1\bm k,n_2\bm k-\bm q})^3}\,,\notag\\
(\Omega_i)_{n1n3}^{n_2}(\bm q)&=\epsilon_{ii_1i_2}\frac{\left(\frac{\partial \hat{H}}{\partial u_{i_1,\bm q}}\right)_{n_1\bm k,n_2\bm k-\bm q}\left(\frac{\partial \hat{H}}{\partial u_{i_2,\bm q}^\star}\right)_{n_2\bm k-\bm q,n_3\bm k}}{(\varepsilon_{n_1,\bm k}-\varepsilon_{n_2,\bm k-\bm q})(\varepsilon_{n_3,\bm k}-\varepsilon_{n_2,\bm k-\bm q})},
\end{align}
$G_2=G_1 \Delta \varepsilon_{n_3\bm k, n_2 \bm k-\bm q}/\Delta \varepsilon_{n_1\bm k, n_2 \bm k-\bm q}$, $\Delta f_{n_1\bm k,n_2\bm k-\bm q} =f_{n_1,\bm k}-f_{n_2,\bm k-\bm q}$, $ \Delta \varepsilon_{n_1\bm k,n_2\bm k-\bm q}=\varepsilon_{n_1,\bm k}-\varepsilon_{n_2,\bm k-\bm q}$, $(\partial \hat{H}/\partial u_{i_1,\bm q})_{n_1\bm k,n_2\bm k-\bm q}=\langle n_1\bm k|\partial \hat{H}/\partial u_{i_1,\bm q}|n_2\bm k-\bm q\rangle$, and $T(\bm q)={\rm Im}\sum_{n_3\neq n_1}(\Omega_i)_{n1n3}^{n_2}(\bm q)(A_{j_1})_{n_3\bm k,n_1\bm k}$ with $(A_{j_1})_{n_3\bm k,n_1\bm k}=\langle n_3\bm k|i\partial_{j_2}|n_1\bm k\rangle$ being the interband Berry connection.

Equation~\eqref{eq_na} is our main result and it does not have any counterpart in the adiabatic perturbation theory. It has several striking features. First, it is closely related to the relaxation processes through the prefactor $\delta_{n_1}^{n_3}$. Interestingly, it relies not on the absolute strength of the relaxation but on the relative ratio of the electronic and phononic lifetime. In the limit that $\eta_p\gg \eta_{n_i}$, $\delta_{n_1}^{n_3}\rightarrow 0$, so that $\lambda_{ij}^{NA}\rightarrow 0$, and only $\lambda_{ij}^A$ contributes. The non-adiabatic correction emerges with $\eta_p$ either comparable with or dominated by certain $\eta_{n_i}$.

This observation sheds light on the nature of adiabaticity. On one hand, for $\lambda_{ij}^{A}$, the correction to the adiabatic result starts at the order of $\omega/\varepsilon_0$, which can be safely ignored if the phonon dynamics is much slower than the electron dynamics, so that the electron motion can follow the atomic one. On the other hand, the life time or the energy exchange rate of electrons shall be much slower than that of phonons, so that the distribution of electronic state is unaffected. Therefore, there are two conditions on the adiabaticity instead of the dynamic condition only. When the second condition is broken, $\lambda_{ij}^{NA}$ emerges.

\begin{figure}[t]
	\includegraphics[width=8.5cm,angle=0]{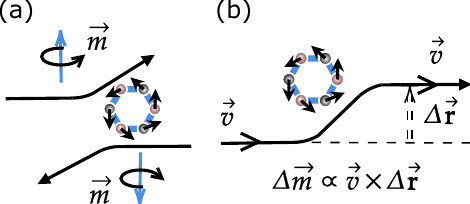}
	\caption{Extrinsic contributions to the phonon magnetic moment. (a) Skew-scattering mechanism. (b) Side-jump Mechanism.}
	\label{fig_fig1}
\end{figure}

Secondly, Eq.~\eqref{eq_na} depends on an essential geometric quantity $ (\Omega_i)_{n1n3}^{n_2}(\bm q)$, which is the interband molecular Berry curvature. For the diagonal element with $n_1=n_3$, $\sum_{n_2\neq n_1}(\Omega_i)_{n1n1}^{n_2}(\bm q)$ recovers the full molecular Berry curvature  introduced previously~\cite{Saito2019,Ren2021,Saparov2022}. Specifically, for two-band systems, $(\Omega_i)_{n1n1}^{n_2}(\bm q)$ coincides with molecular Berry curvature. 

The physical meaning of Eq.~\eqref{eq_na} can be well understood with the help of the modern theory of electronic orbital magnetization $\bm M$~\cite{Xiao2005,Thonhauser2005,Shi2007,Xiao2010}. We start from the following expression of $\bm M$:
\begin{align}\label{eq_mag}
\bm M=\int \frac{d\bm k}{8\pi^3} (\bm m_{n\bm k} f_{n\bm k}+\bm v_{n\bm k}\times \bm A_{n\bm k, n\bm k} f_{n\bm k})\,.
\end{align}
The first terms in both Eq.~\eqref{eq_na} and Eq.~\eqref{eq_mag} contain the orbital magnetic moment $\bm m$. Therefore, the first term in Eq.~\eqref{eq_na} originates from the correction to the distribution function due to the electron-phonon coupling. The appearance of the band-resolved Berry curvature is thus natural, as the transition rate induced by a circular phonon mode is proportional to $[u(\omega,\bm q)\times u^\star(\omega,\bm q)]\cdot (\Omega_i)_{n1n1}^{n_2}(\bm q)$. 

More precisely, the first term in Eq.~\eqref{eq_na} represents a skew-scattering contribution to the phonon magnetic moment.  To see this, we note that in non-magnetic materials, there is equal number of states with positive and negative magnetic moment. Typical examples are transitional metal dichalcogenides, where $\bm m$ takes opposite signs in $K$ and $K^\prime$ valley.  On the other hand, the scattering of those electrons by chiral phonons depends on the band-resolved molecular Berry curvature, which shares the same symmetry property with the magnetic moment. Therefore, states with positive and negative magnetic moment are scattered differently by the chiral phonon, as illustrated in Fig.~\ref{fig_fig1}(a), leading to a net magnetic moment given by the first term in Eq.~\eqref{eq_na}. This is in close analogy to the skew-scattering mechanism of the anomalous Hall effect~\cite{Nagaosa2010}, in which electrons with opposite spin polarization are scattered differently by impurities. Such feature holds in general since both $\bm m$ and molecular Berry curvature are pesudovectors and hence odd funcions of momentum~\cite{suppl}.

By similar logic, the second term in Eq.~\eqref{eq_na} is a side-jump contribution to the phonon magnetic moment. First, it shares a similar structure with the second term in Eq.~\eqref{eq_mag}. We then observe that $G_2(\omega,\bm q)T(\bm q)$ contains the off-diagonal part of both the electron-phonon coupling~(i.e., $u(\omega,\bm q)\times u^\star (\omega, \bm q)\cdot \Omega_{n1n3}^{n2}(\bm q)$) and the position operator~(i.e., $\bm A_{n_3\bm k,n_1\bm k}$). It thus represents the shift of the electronic coordinate during the electron-chiral-phonon coupling. If such coordinate shift is perpendicular to the group velocity of the electron, a net magnetic moment can then emerge, as illustrated by Fig.~\ref{fig_fig1}(b). This closely resembles the side-jump mechanism of the anomalous Hall effect~\cite{Nagaosa2010,Sinitsyn2006}, where electrons change its center-of-mass position after scattered by impurities. 

Equation~\eqref{eq_na} suggests a fundamental scaling relation for the Raman coefficient, in a similar way that the extrinsic contributions to the electronic anomalous Hall effect induces a well-studied scaling relation~\cite{Nagaosa2010,Hou2015}. In metals only the relaxation time of the conducting electron should be considered. In this case, the Raman coefficient satisfies the following scaling relation:
\begin{align}\label{eq_scale}
\lambda_{ij}=\lambda_{ij}^0+\lambda_{ij}^1 \frac{\kappa_{ph}}{\sigma_{el}}\,,
\end{align}
where $\kappa_{ph}$ is the phonon thermal conductivity, $\sigma_{el}$ is the electron conductivity, $\lambda_{ij}^0$ is the intrinsic contribution from the molecular Berry curvature, and $\lambda_{ij}^1$ originates from the skew-scattering and side-jump contribution. The meaning of this scaling relation is two-fold. On one hand, since the phonon magnetic moment can affect many phonon-related phomena, such as the Raman spectrum~\cite{Tang2024,Mustafa2025}, phonon Hall effect~\cite{Strohm2005,Sheng2006,Inyushkin2007,Kagan2008,Wang2009,Zhang2010,Juraschek2022} and so on~\cite{Zhu2018,He2020}, they should also satisfy a similar scaling relation. On the other hand, Eq.~\eqref{eq_scale} suggests that by changing the ratio $\kappa_{ph}/{\sigma_{el}}$, the non-adiabaticity and hence the phonon magnetic moment can be tuned. Generally speaking, $\kappa_{ph}$ and $\sigma_{el}$ respond differently to controllable parameters, such as temperature~\cite{Vallee1994,Debernardi1998,Schultes2013} and doping~\cite{Vignaud2007,Wu2012,Katsiaounis2023}. For example, by changing temperature, the electron relaxation time can have an order-of-magnitude change~\cite{Ashcroft2012,Vignaud2007,Schultes2013}, while the phonon relaxation time change moderately~\cite{Vallee1994,Debernardi1998}. The phonon magnetic moment is then tuned.  
%Finally, we comment that the geometric origin of $\lambda_{ij}^{NA}$ may be more clearly demonstrated when the system can be captured by a two-band model. For example, at $\Gamma$ point, we find that 
%\begin{align}\label{eq_2band}
%	\lambda_{ij}^{NA} & = \beta K_{\alpha \gamma} (g_{0})_{\beta l}
%\end{align}
%where $\left(g_{0}\right)_{\beta l}=Re[\braket{u_{v}|\di \partial_{k_{\beta}}u_{c}} \braket{u_{c}|\di \partial_{k_{l}}u_{v}}]$ is the quantum metric tensor in the electronic valence band, $\beta=\frac{8 \tilde{\beta}^{2}}{a_{0}^{4}} \epsilon_{l \gamma i} \epsilon_{i_{1} i_{2} j} \epsilon_{\alpha i_{1}} \epsilon_{\beta i_{2}}$. $K_{\alpha \gamma}=\left[\delta_{v}^{c} (\hat{v}_{\gamma})_{v} (\hat{v}_{\alpha})_{c} - \delta_{c}^{v} (\hat{v}_{\gamma})_{c} (\hat{v}_{\alpha})_{v}\right] \frac{\Delta \epsilon_{cv} }{ \left( -\omega + \Delta \epsilon_{cv} \right)^{3}}$. $\hat{v}$ is the velocity operator of electrons. $c$ and $v$ are band index for conduction and valence band respectively. Equation~\eqref{eq_2band} reveals the geometric origin of the non-adiabatic phonon magnetic moment at $\Gamma$ point.

%\onecolumngrid\

%\twocolumngrid\

\begin{figure}[t]
	\includegraphics[scale=0.28]{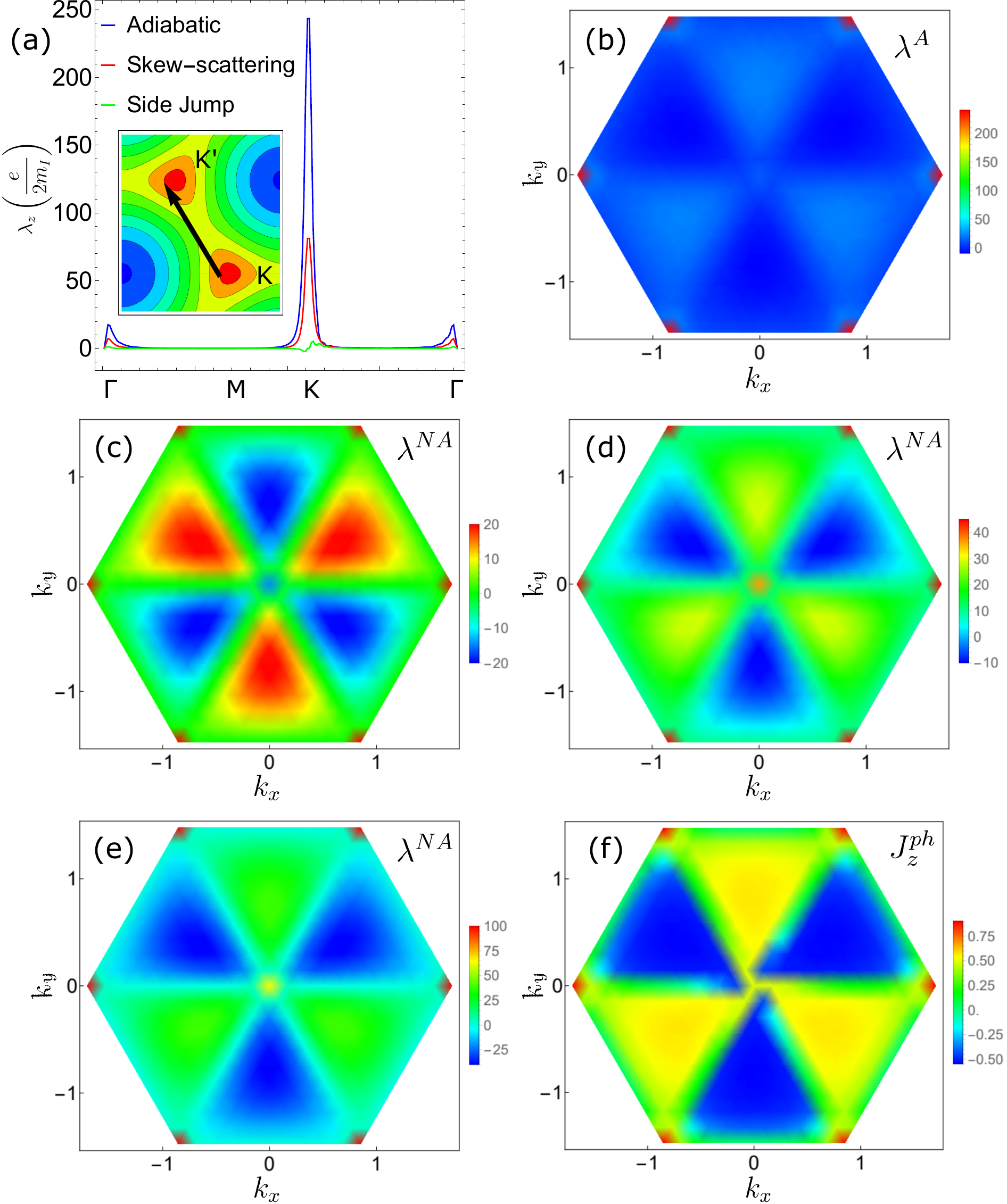}
	\caption{Adiabatic and non-adiabatic contributions to the phonon magnetic moment. (a) The adiabatic, skew-scattering and side-jump contributions along the high-symmetry path in the phonon Brillouin zone. Parameters are $\eta_{p}=0.04$ eV, $\eta_{c}=0.04$ eV, $\eta_{v}=0.04$ eV, and the inset displays the scatter of electrons from $K$ to $K^\prime$ valley by exchanging a phonon at $K$. (b)-(e) $\lambda_{zz}^{A}$ and $\lambda_{zz}^{NA}$ for the highest phonon branch in phonon Brillouin zone with (b) $\eta_{p}=0.02$ eV, $\eta_{c}=0.002$ eV, $\eta_{v}=0.004$ eV; (c) $\eta_{p}=0.02$ eV, $\eta_{c}=0.02$ eV, $\eta_{v}=0.004$ eV; (d) $\eta_{p}=0.02$ eV, $\eta_{c}=0.004$ eV, $\eta_{v}=0.02$ eV; (e) $\eta_{p}=0.004$ eV, $\eta_{c}=0.02$ eV, $\eta_{v}=0.04$ eV, respectively. (f) The angular momentum of phonon in phonon Brillouin zone with the same parameter of (e).}
	\label{fig_fig2}
\end{figure}

\textit{Lattice model.---} As a concrete example, we consider phonons in a honeycomb lattice. The electronic Hamiltonian contains the nearest-neighbour hopping and staggered potential. The electron-phonon coupling is added by allowing the hopping paramter $t$ varies with the bond length~\cite{Ando2006,Balents2019}. In the phonon Hamiltonian, up to third nearest-neighbour elastic potentials are considered. We also use real unit to make it clear that the phonon part has a much smaller energy scale than the electronic part. Detailed model description and parameter choices can be found in supplementary materials~\cite{suppl}. 

We use Eq.~\eqref{eq_lam2} and \eqref{eq_na} to calculate $\lambda_{ij}$, the skew-scattering and side jump contributions to the Raman coefficient, respectively, as shown in Fig.~\ref{fig_fig2}(a). We find that the side jump contribution almost vanishes, as it requires the breaking of particle-hole symmetry for two-band systems, which is preserved in our model~\cite{suppl}. In comparison, both the adiabatic and skew-scattering part of the phonon magnetic moment are peaked near the $\Gamma$ and $K$ piont, and that they can be generally comparable. Such peak is due to the degeneracy of the part of the electronic band around K and $K^\prime$ point as they are connected by phonon momentum at $K$ point. By expanding the lattice Hamitonian around $K$ point, we can simiplify the skew-scattering contribution which clearly shows its geometric origin~\cite{suppl}
\begin{align}
	\lambda_{zz,\bm K}^{SS,NA} = -\frac{1}{2} \left(\delta^{v}_{c}+\delta^{c}_{v}\right) m^{z}_{0} \, \Omega^{z}_{0} G_0\,,
\end{align}
where $G_0=1/(-\omega+2\varepsilon_0)-(\omega\rightarrow -\omega)$, $\Omega^{z}_{0}=\frac{4\beta^{2}v_{f}^{2}}{a_{0}^{4}} \frac{\Delta}{\varepsilon_{0}}$ is the molecular Berry curvature, $\varepsilon_{0}$ is the energy of conduction band, $v_{f}$ is the Fermi velocity, $a_{0}$ is the atomic spacing and $\beta=-\pd{ln t_{0}}{ln a_{0}}$.

In this two-band model, we can also calculate the intrinsic contribution to the Raman coefficient from the molecular Berry curvature. Its relative strength compared with the extrinsic contribution can be described by the ratio: $\lambda_{zz,K}^{SS}/\lambda_{zz,K}^{int}=(\delta_c^v+\delta_v^c)/2$, which reduces to $\delta_v^c/2$ with $n$-type conducting band. This is consistent with the previous scaling relation and shows that to make the extrinsic contribution dominates one has to be far from the adiabatic condition so that $\delta_v^c/2\gg 1$. Since the phonon relaxation time is usually on the order of $ps$~\cite{Vallee1994,Debernardi1998}, while the electron one can change from $10fs$ to $1ns$~\cite{Ashcroft2012,Vignaud2007}, such condition can be reached.

%Finally, we comment that the geometric origin of $\lambda_{ij}^{NA}$ may be more clearly demonstrated when the system can be captured by a two-band model. For example, at $\Gamma$ point, we find that 
%\begin{align}\label{eq_2band}
%	\lambda_{ij}^{NA} & = \beta K_{\alpha \gamma} (g_{0})_{\beta l}
%\end{align}
%where $\left(g_{0}\right)_{\beta l}=Re[\braket{u_{v}|\di \partial_{k_{\beta}}u_{c}} \braket{u_{c}|\di \partial_{k_{l}}u_{v}}]$ is the quantum metric tensor in the electronic valence band, $\beta=\frac{8 \tilde{\beta}^{2}}{a_{0}^{4}} \epsilon_{l \gamma i} \epsilon_{i_{1} i_{2} j} \epsilon_{\alpha i_{1}} \epsilon_{\beta i_{2}}$. $K_{\alpha \gamma}=\left[\delta_{v}^{c} (\hat{v}_{\gamma})_{v} (\hat{v}_{\alpha})_{c} - \delta_{c}^{v} (\hat{v}_{\gamma})_{c} (\hat{v}_{\alpha})_{v}\right] \frac{\Delta \epsilon_{cv} }{ \left( -\omega + \Delta \epsilon_{cv} \right)^{3}}$. $\hat{v}$ is the velocity operator of electrons. $c$ and $v$ are band index for conduction and valence band respectively. Equation~\eqref{eq_2band} reveals the geometric origin of the non-adiabatic phonon magnetic moment at $\Gamma$ point.

As different contributions to the phonon magnetic moment are sensitive to the relaxation processes, we plot the distribution of $\lambda_{zz}$ across the whole phonon Brillouin zone in different regimes in Fig.~\ref{fig_fig2}(b)-(e). We choose the phonon band with the highest energy. In Fig.~\ref{fig_fig2}(b), $\eta_p\gg \eta_c,\eta_v$, and the adiabaticity is restored. In this case, the adiabatic contribution is order-of-magnitude larger than the non-adiabatic one. The phonon magnetic moment is highly non-uniform, with a large peak at two valleys and small variation in other regions. 

In Fig.~\ref{fig_fig2}(c-d), the parameters are so chosen that only electrons in one band is adiabatic and those in the other one is not. The non-adiabatic phonon magnetic moment shows a $C_3$ symmetry, consistent with the point-group symmetry of the lattice. Moreover, although it is peaked at the valley, there are obvious variations in other regions of the Brillouin zone, even with sign changing. Due to the asymmetry between the valence and conduction band parameters, both the skew-scattering and side jump mechanism contribute, and they should correspond to the common part and the difference between (c) and (d), respectively. In Fig.~\ref{fig_fig2}(e), both bands are in the non-adiabatic regime and the resulting non-adiabatic magnetic moment increases.

In Fig.~\ref{fig_fig2}(f), we show the phonon angular momentum with the electronic conduction band in the adiabatic regime. This can be realized with hole-doping in the electronic degree of freedom. We find that since the phonon magnetic moment is highly nonuniform in the Brillouin zone, so is the angular momentum.  Moreover, the angular momentum at different regions are roughly on the same order, showing the importance of the non-adiabatic contribution, as the adiabatic magnetic moment is mainly localized near the $K$ and $K^\prime$ point. This inhomogeneity can manifest in the optical response of phonons.

\begin{acknowledgements}
Q. N. is supported  the National Key R${\rm \&}$D Program under grant Nos. 2023YFA1406300. Y. G. and Q. N. are supported by the National Natural Science Foundation of China (Grants Nos. 12234017, 12374164). R. Xue and Z. Qiao are supported by the National Natural Science Foundation of China (Grants Nos. 12474158  and 12488101), Anhui Initiative in Quantum Information Technologies (AHY170000). We also thank the support from the Innovation Program for Quantum Science and Technology (2021ZD0302800) and the Supercomputing Center of University of Science and Technology of China.
\end{acknowledgements}

\bibliography{PmBib}

@misc{suppl,
  note = "See Supplemental Material [url] for details of the derivation and lattice model, which includes Refs. [64-66]."
}

@Article{Zhang2014,
  author    = {Zhang, Lifa and Niu, Qian},
  journal   = {Physical Review Letters},
  title     = {Angular Momentum of Phonons and the Einstein–de Haas Effect},
  year      = {2014},
  month     = feb,
  number    = {8},
  pages     = {085503},
  volume    = {112},
  doi       = {10.1103/physrevlett.112.085503},
  publisher = {American Physical Society (APS)},
}

@Article{Ueda2023,
  author    = {Ueda, Hiroki and García-Fernández, Mirian and Agrestini, Stefano and Romao, Carl P. and van den Brink, Jeroen and Spaldin, Nicola A. and Zhou, Ke-Jin and Staub, Urs},
  journal   = {Nature},
  title     = {Chiral phonons in quartz probed by X-rays},
  year      = {2023},
  month     = jun,
  number    = {7967},
  pages     = {946--950},
  volume    = {618},
  doi       = {10.1038/s41586-023-06016-5},
  publisher = {Springer Science and Business Media LLC},
}

@Article{Chen2020,
  author    = {Chen, Hsiao-Yi and Sangalli, Davide and Bernardi, Marco},
  journal   = {Physical Review Letters},
  title     = {Exciton-Phonon Interaction and Relaxation Times from First Principles},
  year      = {2020},
  month     = aug,
  number    = {10},
  pages     = {107401},
  volume    = {125},
  doi       = {10.1103/physrevlett.125.107401},
  publisher = {American Physical Society (APS)},
}

@Article{Lujan2024,
  author    = {Lujan, David and Choe, Jeongheon and Chaudhary, Swati and Ye, Gaihua and Nnokwe, Cynthia and Rodriguez-Vega, Martin and He, Jiaming and Gao, Frank Y. and Nunley, T. Nathan and Baldini, Edoardo and Zhou, Jianshi and Fiete, Gregory A. and He, Rui and Li, Xiaoqin},
  journal   = {Proceedings of the National Academy of Sciences},
  title     = {Spin–orbit exciton–induced phonon chirality in a quantum magnet},
  year      = {2024},
  month     = mar,
  number    = {11},
  pages     = {e2304360121},
  volume    = {121},
  doi       = {10.1073/pnas.2304360121},
  publisher = {Proceedings of the National Academy of Sciences},
}

@Article{He2020,
  author    = {He, Minhao and Rivera, Pasqual and Van Tuan, Dinh and Wilson, Nathan P. and Yang, Min and Taniguchi, Takashi and Watanabe, Kenji and Yan, Jiaqiang and Mandrus, David G. and Yu, Hongyi and Dery, Hanan and Yao, Wang and Xu, Xiaodong},
  journal   = {Nature Communications},
  title     = {Valley phonons and exciton complexes in a monolayer semiconductor},
  year      = {2020},
  month     = jan,
  number    = {1},
  pages     = {618},
  volume    = {11},
  doi       = {10.1038/s41467-020-14472-0},
  publisher = {Springer Science and Business Media LLC},
}

@Article{Gao2023,
  author    = {Gao, Yi and Pan, Yang and Zhou, Jun and Zhang, Lifa},
  journal   = {Physical Review B},
  title     = {Chiral phonon mediated high-temperature superconductivity},
  year      = {2023},
  month     = aug,
  number    = {6},
  pages     = {064510},
  volume    = {108},
  doi       = {10.1103/physrevb.108.064510},
  publisher = {American Physical Society (APS)},
}

@Article{Zhang2019,
  author    = {Zhang, Xiaoou and Zhang, Yinhan and Okamoto, Satoshi and Xiao, Di},
  journal   = {Physical Review Letters},
  title     = {Thermal Hall Effect Induced by Magnon-Phonon Interactions},
  year      = {2019},
  month     = oct,
  number    = {16},
  pages     = {167202},
  volume    = {123},
  doi       = {10.1103/physrevlett.123.167202},
  publisher = {American Physical Society (APS)},
}

@Article{Zhang2010,
  author    = {Zhang, Lifa and Ren, Jie and Wang, Jian-Sheng and Li, Baowen},
  journal   = {Physical Review Letters},
  title     = {Topological Nature of the Phonon Hall Effect},
  year      = {2010},
  month     = nov,
  number    = {22},
  pages     = {225901},
  volume    = {105},
  doi       = {10.1103/physrevlett.105.225901},
  publisher = {American Physical Society (APS)},
}

@Article{Saito2019,
  author    = {Saito, Takuma and Misaki, Kou and Ishizuka, Hiroaki and Nagaosa, Naoto},
  journal   = {Physical Review Letters},
  title     = {Berry Phase of Phonons and Thermal Hall Effect in Nonmagnetic Insulators},
  year      = {2019},
  month     = dec,
  number    = {25},
  pages     = {255901},
  volume    = {123},
  doi       = {10.1103/physrevlett.123.255901},
  publisher = {American Physical Society (APS)},
}

@Article{Ren2021,
  author    = {Ren, Yafei and Xiao, Cong and Saparov, Daniyar and Niu, Qian},
  journal   = {Physical Review Letters},
  title     = {Phonon Magnetic Moment from Electronic Topological Magnetization},
  year      = {2021},
  month     = oct,
  number    = {18},
  pages     = {186403},
  volume    = {127},
  doi       = {10.1103/physrevlett.127.186403},
  publisher = {American Physical Society (APS)},
}

@Article{Hu2021,
  author    = {Hu, Lun-Hui and Yu, Jiabin and Garate, Ion and Liu, Chao-Xing},
  journal   = {Physical Review Letters},
  title     = {Phonon Helicity Induced by Electronic Berry Curvature in Dirac Materials},
  year      = {2021},
  month     = sep,
  number    = {12},
  pages     = {125901},
  volume    = {127},
  doi       = {10.1103/physrevlett.127.125901},
  publisher = {American Physical Society (APS)},
}

@Article{Saparov2022,
  author    = {Saparov, Daniyar and Xiong, Bangguo and Ren, Yafei and Niu, Qian},
  journal   = {Physical Review B},
  title     = {Lattice dynamics with molecular Berry curvature: Chiral optical phonons},
  year      = {2022},
  month     = feb,
  number    = {6},
  pages     = {064303},
  volume    = {105},
  doi       = {10.1103/physrevb.105.064303},
  publisher = {American Physical Society (APS)},
}

@Article{Strohm2005,
  author    = {Strohm, C. and Rikken, G. L. J. A. and Wyder, P.},
  journal   = {Physical Review Letters},
  title     = {Phenomenological Evidence for the Phonon Hall Effect},
  year      = {2005},
  month     = oct,
  number    = {15},
  pages     = {155901},
  volume    = {95},
  doi       = {10.1103/physrevlett.95.155901},
  publisher = {American Physical Society (APS)},
}

@Article{Sheng2006,
  author    = {Sheng, L. and Sheng, D. N. and Ting, C. S.},
  journal   = {Physical Review Letters},
  title     = {Theory of the Phonon Hall Effect in Paramagnetic Dielectrics},
  year      = {2006},
  month     = apr,
  number    = {15},
  pages     = {155901},
  volume    = {96},
  doi       = {10.1103/physrevlett.96.155901},
  publisher = {American Physical Society (APS)},
}

@Article{Inyushkin2007,
  author    = {Inyushkin, A. V. and Taldenkov, A. N.},
  journal   = {JETP Letters},
  title     = {On the phonon Hall effect in a paramagnetic dielectric},
  year      = {2007},
  month     = nov,
  number    = {6},
  pages     = {379--382},
  volume    = {86},
  doi       = {10.1134/s0021364007180075},
  publisher = {Pleiades Publishing Ltd},
}

@Article{Kagan2008,
  author    = {Kagan, Yu. and Maksimov, L. A.},
  journal   = {Physical Review Letters},
  title     = {Anomalous Hall Effect for the Phonon Heat Conductivity in Paramagnetic Dielectrics},
  year      = {2008},
  month     = apr,
  number    = {14},
  pages     = {145902},
  volume    = {100},
  doi       = {10.1103/physrevlett.100.145902},
  publisher = {American Physical Society (APS)},
}

@Article{Wang2009,
  author    = {Wang, Jian-Sheng and Zhang, Lifa},
  journal   = {Physical Review B},
  title     = {Phonon Hall thermal conductivity from the Green-Kubo formula},
  year      = {2009},
  month     = jul,
  number    = {1},
  pages     = {012301},
  volume    = {80},
  doi       = {10.1103/physrevb.80.012301},
  publisher = {American Physical Society (APS)},
}

@Article{Wang2024,
  author    = {Wang, Qian and Long, Meng-Qiu and Wang, Yun-Peng},
  journal   = {Physical Review B},
  title     = {Magnetic moments of chiral phonons induced by coupling with magnons},
  year      = {2024},
  month     = jul,
  number    = {2},
  pages     = {024423},
  volume    = {110},
  doi       = {10.1103/physrevb.110.024423},
  publisher = {American Physical Society (APS)},
}

@Article{Cheng2020,
  author    = {Cheng, Bing and Schumann, T. and Wang, Youcheng and Zhang, X. and Barbalas, D. and Stemmer, S. and Armitage, N. P.},
  journal   = {Nano Letters},
  title     = {A Large Effective Phonon Magnetic Moment in a Dirac Semimetal},
  year      = {2020},
  month     = jul,
  number    = {8},
  pages     = {5991--5996},
  volume    = {20},
  doi       = {10.1021/acs.nanolett.0c01983},
  publisher = {American Chemical Society (ACS)},
}

@Article{Bistoni2021,
  author    = {Bistoni, Oliviero and Mauri, Francesco and Calandra, Matteo},
  journal   = {Physical Review Letters},
  title     = {Intrinsic Vibrational Angular Momentum from Nonadiabatic Effects in Noncollinear Magnetic Molecules},
  year      = {2021},
  month     = jun,
  number    = {22},
  pages     = {225703},
  volume    = {126},
  doi       = {10.1103/physrevlett.126.225703},
  publisher = {American Physical Society (APS)},
}

@Article{Bonini2023,
  author    = {Bonini, John and Ren, Shang and Vanderbilt, David and Stengel, Massimiliano and Dreyer, Cyrus E. and Coh, Sinisa},
  journal   = {Physical Review Letters},
  title     = {Frequency Splitting of Chiral Phonons from Broken Time-Reversal Symmetry in CrI3},
  year      = {2023},
  month     = feb,
  number    = {8},
  pages     = {086701},
  volume    = {130},
  doi       = {10.1103/physrevlett.130.086701},
  publisher = {American Physical Society (APS)},
}

@Article{Juraschek2022,
  author    = {Juraschek, Dominik M. and Neuman, Tomáš and Narang, Prineha},
  journal   = {Physical Review Research},
  title     = {Giant effective magnetic fields from optically driven chiral phonons in 4f paramagnets},
  year      = {2022},
  month     = feb,
  number    = {1},
  pages     = {013129},
  volume    = {4},
  doi       = {10.1103/physrevresearch.4.013129},
  publisher = {American Physical Society (APS)},
}

@Article{Nomura2019,
  author    = {Nomura, T. and Zhang, X.-X. and Zherlitsyn, S. and Wosnitza, J. and Tokura, Y. and Nagaosa, N. and Seki, S.},
  journal   = {Physical Review Letters},
  title     = {Phonon Magnetochiral Effect},
  year      = {2019},
  month     = apr,
  number    = {14},
  pages     = {145901},
  volume    = {122},
  doi       = {10.1103/physrevlett.122.145901},
  publisher = {American Physical Society (APS)},
}

@Article{Zabalo2022,
  author    = {Zabalo, Asier and Dreyer, Cyrus E. and Stengel, Massimiliano},
  journal   = {Physical Review B},
  title     = {Rotational g factors and Lorentz forces of molecules and solids from density functional perturbation theory},
  year      = {2022},
  month     = mar,
  number    = {9},
  pages     = {094305},
  volume    = {105},
  doi       = {10.1103/physrevb.105.094305},
  publisher = {American Physical Society (APS)},
}

@Article{Xiao2005,
  author    = {Xiao, Di and Shi, Junren and Niu, Qian},
  journal   = {Physical Review Letters},
  title     = {Berry Phase Correction to Electron Density of States in Solids},
  year      = {2005},
  month     = sep,
  number    = {13},
  pages     = {137204},
  volume    = {95},
  doi       = {10.1103/physrevlett.95.137204},
  publisher = {American Physical Society (APS)},
}

@Article{Shi2007,
  author    = {Shi, Junren and Vignale, G. and Xiao, Di and Niu, Qian},
  journal   = {Physical Review Letters},
  title     = {Quantum Theory of Orbital Magnetization and Its Generalization to Interacting Systems},
  year      = {2007},
  month     = nov,
  number    = {19},
  pages     = {197202},
  volume    = {99},
  doi       = {10.1103/physrevlett.99.197202},
  publisher = {American Physical Society (APS)},
}

@Article{Xiao2010,
  author    = {Xiao, Di and Chang, Ming-Che and Niu, Qian},
  journal   = {Reviews of Modern Physics},
  title     = {Berry phase effects on electronic properties},
  year      = {2010},
  month     = jul,
  number    = {3},
  pages     = {1959--2007},
  volume    = {82},
  doi       = {10.1103/revmodphys.82.1959},
  publisher = {American Physical Society (APS)},
}

@Article{Kozii2024,
  author    = {Kozii, Vladyslav and Fu, Liang},
  journal   = {Physical Review B},
  title     = {Non-Hermitian topological theory of finite-lifetime quasiparticles: Prediction of bulk Fermi arc due to exceptional point},
  year      = {2024},
  month     = jun,
  number    = {23},
  pages     = {235139},
  volume    = {109},
  doi       = {10.1103/physrevb.109.235139},
  publisher = {American Physical Society (APS)},
}

@Article{Thonhauser2005,
  author    = {Thonhauser, T. and Ceresoli, Davide and Vanderbilt, David and Resta, R.},
  journal   = {Physical Review Letters},
  title     = {Orbital Magnetization in Periodic Insulators},
  year      = {2005},
  month     = sep,
  number    = {13},
  pages     = {137205},
  volume    = {95},
  doi       = {10.1103/physrevlett.95.137205},
  publisher = {American Physical Society (APS)},
}

@Article{Geng2023,
  author    = {Geng, H. and Wei, J. Y. and Zou, M. H. and Sheng, L. and Chen, Wei and Xing, D. Y.},
  journal   = {Physical Review B},
  title     = {Nonreciprocal charge and spin transport induced by non-Hermitian skin effect in mesoscopic heterojunctions},
  year      = {2023},
  month     = jan,
  number    = {3},
  pages     = {035306},
  volume    = {107},
  doi       = {10.1103/physrevb.107.035306},
  publisher = {American Physical Society (APS)},
}

@Article{Wang2022,
  author    = {Wang, Jiong-Hao and Tao, Yu-Liang and Xu, Yong},
  journal   = {Chinese Physics Letters},
  title     = {Anomalous Transport Induced by Non-Hermitian Anomalous Berry Connection in Non-Hermitian Systems},
  year      = {2022},
  month     = jan,
  number    = {1},
  pages     = {010301},
  volume    = {39},
  doi       = {10.1088/0256-307x/39/1/010301},
  publisher = {IOP Publishing},
}

@Article{Yang2024,
  author    = {Yang, Guang and Li, Yong-Kang and Fu, Yongxu and Wang, Zhenduo and Zhang, Yi},
  journal   = {Physical Review B},
  title     = {Complex semiclassical theory for non-Hermitian quantum systems},
  year      = {2024},
  month     = jan,
  number    = {4},
  pages     = {045110},
  volume    = {109},
  doi       = {10.1103/physrevb.109.045110},
  publisher = {American Physical Society (APS)},
}

@Article{Nagai2020,
  author    = {Nagai, Yuki and Qi, Yang and Isobe, Hiroki and Kozii, Vladyslav and Fu, Liang},
  journal   = {Physical Review Letters},
  title     = {DMFT Reveals the Non-Hermitian Topology and Fermi Arcs in Heavy-Fermion Systems},
  year      = {2020},
  month     = nov,
  number    = {22},
  pages     = {227204},
  volume    = {125},
  doi       = {10.1103/physrevlett.125.227204},
  publisher = {American Physical Society (APS)},
}

@Article{Yao2018,
  author    = {Yao, Shunyu and Wang, Zhong},
  journal   = {Physical Review Letters},
  title     = {Edge States and Topological Invariants of Non-Hermitian Systems},
  year      = {2018},
  month     = aug,
  number    = {8},
  pages     = {086803},
  volume    = {121},
  doi       = {10.1103/physrevlett.121.086803},
  publisher = {American Physical Society (APS)},
}

@Article{Michishita2020,
  author    = {Michishita, Yoshihiro and Peters, Robert},
  journal   = {Physical Review Letters},
  title     = {Equivalence of Effective Non-Hermitian Hamiltonians in the Context of Open Quantum Systems and Strongly Correlated Electron Systems},
  year      = {2020},
  month     = may,
  number    = {19},
  pages     = {196401},
  volume    = {124},
  doi       = {10.1103/physrevlett.124.196401},
  publisher = {American Physical Society (APS)},
}

@Article{Chen2018,
  author    = {Chen, Yu and Zhai, Hui},
  journal   = {Physical Review B},
  title     = {Hall conductance of a non-Hermitian Chern insulator},
  year      = {2018},
  month     = dec,
  number    = {24},
  pages     = {245130},
  volume    = {98},
  doi       = {10.1103/physrevb.98.245130},
  publisher = {American Physical Society (APS)},
}

@Article{Yokomizo2019,
  author    = {Yokomizo, Kazuki and Murakami, Shuichi},
  journal   = {Physical Review Letters},
  title     = {Non-Bloch Band Theory of Non-Hermitian Systems},
  year      = {2019},
  month     = aug,
  number    = {6},
  pages     = {066404},
  volume    = {123},
  doi       = {10.1103/physrevlett.123.066404},
  publisher = {American Physical Society (APS)},
}

@Article{Pan2020,
  author    = {Pan, Lei and Chen, Xin and Chen, Yu and Zhai, Hui},
  journal   = {Nature Physics},
  title     = {Non-Hermitian linear response theory},
  year      = {2020},
  month     = may,
  number    = {7},
  pages     = {767--771},
  volume    = {16},
  doi       = {10.1038/s41567-020-0889-6},
  publisher = {Springer Science and Business Media LLC},
}

@Article{Shen2018,
  author    = {Shen, Huitao and Zhen, Bo and Fu, Liang},
  journal   = {Physical Review Letters},
  title     = {Topological Band Theory for Non-Hermitian Hamiltonians},
  year      = {2018},
  month     = apr,
  number    = {14},
  pages     = {146402},
  volume    = {120},
  doi       = {10.1103/physrevlett.120.146402},
  publisher = {American Physical Society (APS)},
}

@Article{Kaplan2023,
  author    = {Kaplan, Daniel and Holder, Tobias and Yan, Binghai},
  journal   = {Nature Communications},
  title     = {General nonlinear Hall current in magnetic insulators beyond the quantum anomalous Hall effect},
  year      = {2023},
  month     = may,
  number    = {1},
  pages     = {3053},
  volume    = {14},
  doi       = {10.1038/s41467-023-38734-9},
  publisher = {Springer Science and Business Media LLC},
}

@Article{Kaplan2020,
  author    = {Kaplan, Daniel and Holder, Tobias and Yan, Binghai},
  journal   = {Physical Review Letters},
  title     = {Nonvanishing Subgap Photocurrent as a Probe of Lifetime Effects},
  year      = {2020},
  month     = nov,
  number    = {22},
  pages     = {227401},
  volume    = {125},
  doi       = {10.1103/physrevlett.125.227401},
  publisher = {American Physical Society (APS)},
}

@Article{Nagaosa2010,
  author    = {Nagaosa, Naoto and Sinova, Jairo and Onoda, Shigeki and MacDonald, A. H. and Ong, N. P.},
  journal   = {Reviews of Modern Physics},
  title     = {Anomalous Hall effect},
  year      = {2010},
  month     = may,
  number    = {2},
  pages     = {1539--1592},
  volume    = {82},
  doi       = {10.1103/revmodphys.82.1539},
  publisher = {American Physical Society (APS)},
}

@Article{Sinitsyn2006,
  author    = {Sinitsyn, N. A. and Niu, Q. and MacDonald, A. H.},
  journal   = {Physical Review B},
  title     = {Coordinate shift in the semiclassical Boltzmann equation and the anomalous Hall effect},
  year      = {2006},
  month     = feb,
  number    = {7},
  pages     = {075318},
  volume    = {73},
  doi       = {10.1103/physrevb.73.075318},
  publisher = {American Physical Society (APS)},
}

@Article{Xiao2019,
  author    = {Xiao, Cong and Liu, Yi and Yuan, Zhe and Yang, Shengyuan A. and Niu, Qian},
  journal   = {Physical Review B},
  title     = {Temperature dependence of the side-jump spin Hall conductivity},
  year      = {2019},
  month     = aug,
  number    = {8},
  pages     = {085425},
  volume    = {100},
  doi       = {10.1103/physrevb.100.085425},
  publisher = {American Physical Society (APS)},
}

@Article{Raimondi2011,
  author    = {Raimondi, R. and Schwab, P. and Gorini, C. and Vignale, G.},
  journal   = {Annalen der Physik},
  title     = {Spin‐orbit interaction in a two‐dimensional electron gas: A SU(2) formulation},
  year      = {2011},
  month     = dec,
  number    = {3–4},
  volume    = {524},
  doi       = {10.1002/andp.201100253},
  publisher = {Wiley},
}

@Article{Ando2006,
  author    = {Ando, Tsuneya},
  journal   = {Journal of the Physical Society of Japan},
  title     = {Anomaly of Optical Phonon in Monolayer Graphene},
  year      = {2006},
  month     = dec,
  number    = {12},
  pages     = {124701},
  volume    = {75},
  doi       = {10.1143/jpsj.75.124701},
  publisher = {Physical Society of Japan},
}

@Article{Balents2019,
  author    = {Balents, Leon},
  journal   = {SciPost Physics},
  title     = {General continuum model for twisted bilayer graphene and arbitrary smooth deformations},
  year      = {2019},
  month     = oct,
  number    = {4},
  pages     = {048},
  volume    = {7},
  doi       = {10.21468/scipostphys.7.4.048},
  publisher = {Stichting SciPost},
}

@Article{Zhang2022,
  author    = {Zhang, Tiantian and Murakami, Shuichi},
  journal   = {Physical Review Research},
  title     = {Chiral phonons and pseudoangular momentum in nonsymmorphic systems},
  year      = {2022},
  month     = feb,
  number    = {1},
  pages     = {l012024},
  volume    = {4},
  doi       = {10.1103/physrevresearch.4.l012024},
  publisher = {American Physical Society (APS)},
}

@Article{Zhang2023,
  author    = {Zhang, Tiantian and Huang, Zhiheng and Pan, Zitian and Du, Luojun and Zhang, Guangyu and Murakami, Shuichi},
  journal   = {Nano Letters},
  title     = {Weyl Phonons in Chiral Crystals},
  year      = {2023},
  month     = aug,
  number    = {16},
  pages     = {7561--7567},
  volume    = {23},
  doi       = {10.1021/acs.nanolett.3c02132},
  publisher = {American Chemical Society (ACS)},
}

@Article{Zhang2018,
  author    = {Zhang, Tiantian and Song, Zhida and Alexandradinata, A. and Weng, Hongming and Fang, Chen and Lu, Ling and Fang, Zhong},
  journal   = {Physical Review Letters},
  title     = {Double-Weyl Phonons in Transition-Metal Monosilicides},
  year      = {2018},
  month     = jan,
  number    = {1},
  pages     = {016401},
  volume    = {120},
  doi       = {10.1103/physrevlett.120.016401},
  publisher = {American Physical Society (APS)},
}

@Article{Miao2018,
  author    = {Miao, H. and Zhang, T. T. and Wang, L. and Meyers, D. and Said, A. H. and Wang, Y. L. and Shi, Y. G. and Weng, H. M. and Fang, Z. and Dean, M. P. M.},
  journal   = {Physical Review Letters},
  title     = {Observation of Double Weyl Phonons in Parity-Breaking FeSi},
  year      = {2018},
  month     = jul,
  number    = {3},
  pages     = {035302},
  volume    = {121},
  doi       = {10.1103/physrevlett.121.035302},
  publisher = {American Physical Society (APS)},
}

@Article{Zhang2020,
  author    = {Zhang, Tiantian and Takahashi, Ryo and Fang, Chen and Murakami, Shuichi},
  journal   = {Physical Review B},
  title     = {Twofold quadruple Weyl nodes in chiral cubic crystals},
  year      = {2020},
  month     = sep,
  number    = {12},
  pages     = {125148},
  volume    = {102},
  doi       = {10.1103/physrevb.102.125148},
  publisher = {American Physical Society (APS)},
}

@Article{Li2021,
  author    = {Li, Haoxiang and Zhang, Tiantian and Said, A. and Fu, Y. and Fabbris, G. and Mazzone, D. G. and Zhang, J. and Lapano, J. and Lee, H. N. and Lei, H. C. and Dean, M. P. M. and Murakami, S. and Miao, H.},
  journal   = {Physical Review B},
  title     = {Observation of a chiral wave function in the twofold-degenerate quadruple Weyl system BaPtGe},
  year      = {2021},
  month     = may,
  number    = {18},
  pages     = {184301},
  volume    = {103},
  doi       = {10.1103/physrevb.103.184301},
  publisher = {American Physical Society (APS)},
}

@Article{Zhu2018,
  author    = {Zhu, Hanyu and Yi, Jun and Li, Ming-Yang and Xiao, Jun and Zhang, Lifa and Yang, Chih-Wen and Kaindl, Robert A. and Li, Lain-Jong and Wang, Yuan and Zhang, Xiang},
  journal   = {Science},
  title     = {Observation of chiral phonons},
  year      = {2018},
  month     = feb,
  number    = {6375},
  pages     = {579--582},
  volume    = {359},
  doi       = {10.1126/science.aar2711},
  publisher = {American Association for the Advancement of Science (AAAS)},
}

@Article{Mustafa2025,
  author    = {Mustafa, Hussam and Nnokwe, Cynthia and Ye, Gaihua and Fang, Mengqi and Chaudhary, Swati and Yan, Jia-An and Wu, Kai and Cunningham, Connor J. and Hemesath, Colin M. and Stollenwerk, Andrew James and Shand, Paul M. and Yang, Eui-Hyeok and Fiete, Gregory A. and He, Rui and Jin, Wencan},
  journal   = {ACS Nano},
  title     = {Origin of Large Effective Phonon Magnetic Moments in Monolayer MoS2},
  year      = {2025},
  month     = mar,
  number    = {11},
  pages     = {11241--11248},
  volume    = {19},
  doi       = {10.1021/acsnano.4c18906},
  publisher = {American Chemical Society (ACS)},
}

@Article{Tang2024,
  author    = {Tang, Chunli and Ye, Gaihua and Nnokwe, Cynthia and Fang, Mengqi and Xiang, Li and Mahjouri-Samani, Masoud and Smirnov, Dmitry and Yang, Eui-Hyeok and Wang, Tingting and Zhang, Lifa and He, Rui and Jin, Wencan},
  journal   = {Physical Review B},
  title     = {Exciton-activated effective phonon magnetic moment in monolayer MoS2},
  year      = {2024},
  month     = apr,
  number    = {15},
  pages     = {155426},
  volume    = {109},
  doi       = {10.1103/physrevb.109.155426},
  publisher = {American Physical Society (APS)},
}

@Article{Schultes2013,
  author    = {Schultes, F. J. and Christian, T. and Jones-Albertus, R. and Pickett, E. and Alberi, K. and Fluegel, B. and Liu, T. and Misra, P. and Sukiasyan, A. and Yuen, H. and Haegel, N. M.},
  journal   = {Applied Physics Letters},
  title     = {Temperature dependence of diffusion length, lifetime and minority electron mobility in GaInP},
  year      = {2013},
  month     = dec,
  number    = {24},
  pages     = {242106},
  volume    = {103},
  doi       = {10.1063/1.4847635},
  publisher = {AIP Publishing},
}

@Article{Debernardi1998,
  author    = {Debernardi, Alberto},
  journal   = {Physical Review B},
  title     = {Phonon linewidth in III-V semiconductors from density-functional perturbation theory},
  year      = {1998},
  month     = may,
  number    = {20},
  pages     = {12847--12858},
  volume    = {57},
  doi       = {10.1103/physrevb.57.12847},
  publisher = {American Physical Society (APS)},
}

@Article{Vallee1994,
  author    = {Vallée, Fabrice},
  journal   = {Physical Review B},
  title     = {Time-resolved investigation of coherent LO-phonon relaxation in III-V semiconductors},
  year      = {1994},
  month     = jan,
  number    = {4},
  pages     = {2460--2468},
  volume    = {49},
  doi       = {10.1103/physrevb.49.2460},
  publisher = {American Physical Society (APS)},
}

@Article{Vignaud2007,
  author    = {Vignaud, D. and Yarekha, D. A. and Lampin, J. F. and Zaknoune, M. and Godey, S. and Mollot, F.},
  journal   = {Applied Physics Letters},
  title     = {Electron lifetime measurements of heavily C-doped InGaAs and GaAsSb as a function of the doping density},
  year      = {2007},
  month     = jun,
  number    = {24},
  pages     = {242104},
  volume    = {90},
  doi       = {10.1063/1.2748336},
  publisher = {AIP Publishing},
}

@Article{Wu2012,
  author    = {Wu, Shiwei and Liu, Wei-Tao and Liang, Xiaogan and Schuck, P. James and Wang, Feng and Shen, Y. Ron and Salmeron, Miquel},
  journal   = {Nano Letters},
  title     = {Hot Phonon Dynamics in Graphene},
  year      = {2012},
  month     = nov,
  number    = {11},
  pages     = {5495--5499},
  volume    = {12},
  doi       = {10.1021/nl301997r},
  publisher = {American Chemical Society (ACS)},
}

@Article{Katsiaounis2023,
  author    = {Katsiaounis, Stavros and Delikoukos, Nikos and Michail, Antonios and Parthenios, John and Papagelis, Konstantinos},
  journal   = {Carbon},
  title     = {G phonon linewidth and phonon-phonon interaction in p-type doped CVD graphene crystals},
  year      = {2023},
  month     = nov,
  pages     = {118449},
  volume    = {215},
  doi       = {10.1016/j.carbon.2023.118449},
  publisher = {Elsevier BV},
}

@Article{Hou2015,
  author    = {Hou, Dazhi and Su, Gang and Tian, Yuan and Jin, Xiaofeng and Yang, Shengyuan A. and Niu, Qian},
  journal   = {Physical Review Letters},
  title     = {Multivariable Scaling for the Anomalous Hall Effect},
  year      = {2015},
  month     = may,
  number    = {21},
  pages     = {217203},
  volume    = {114},
  doi       = {10.1103/physrevlett.114.217203},
  publisher = {American Physical Society (APS)},
}

@Article{Yu2024,
  author    = {Yu, Tao and Zou, Ji and Zeng, Bowen and Rao, J.W. and Xia, Ke},
  journal   = {Physics Reports},
  title     = {Non-Hermitian topological magnonics},
  year      = {2024},
  month     = apr,
  pages     = {1--86},
  volume    = {1062},
  doi       = {10.1016/j.physrep.2024.01.006},
  publisher = {Elsevier BV},
}

@Book{Ashcroft2012,
  author    = {Ashcroft, Neil W. and Mermin, N. David},
  publisher = {Brooks/Cole Thomson Learning},
  title     = {Solid state physics},
  year      = {2012},
  edition   = {Repr.},
  isbn      = {9780030839931},
  pagetotal = {826},
  ppn_gvk   = {756648025},
}

\end{document}